# Chip-based Brillouin processing for carrier recovery in coherent optical communications

ELIAS GIACOUMIDIS,[1,2,\*,#] AMOL CHOUDHARY, [1,2,\*,#] ERIC MAGI,[1,2] DAVID MARPAUNG,[1,2] KHU VU,[3] PAN MA,[3] DUK-YONG CHOI,[3] STEVE MADDEN,[3] BILL CORCORAN,[4] MARK PELUSI,[1,2] AND BENJAMIN J. EGGLETON,[1,2\*]

[1]*Institute of Photonics and Optical Science (IPOS), School of Physics, University of Sydney, Sydney, Australia.*
[2]*The University of Sydney Nano Institute, University of Sydney, Sydney, Australia.*
[3]*Laser Physics Centre, Australian National University, Canberra, Australia.*
[4]*Monash University, Melbourne, Victoria, Australia.*

[#]These authors contributed equally to this work. E. Giacoumidis is now with Dublin City University, Ireland.
[\*]Corresponding authors: eliasgiacoumidis@hotmail.com, amol.choudhary@sydney.edu.au, egg@physics.usyd.edu.au

**Modern fiber-optic coherent communications employ advanced, spectrally-efficient modulation formats that require sophisticated narrow linewidth local oscillators (LOs) and complex digital signal processing (DSP). Self-coherent optical orthogonal frequency-division multiplexing (Self-CO-OFDM) is a modern technology that retrieves the frequency and phase information from the extracted carrier without employing a LO or additional DSP. However, a wide guardband is typically required to easily filter out the optical carrier at the receiver, thus discarding many OFDM middle subcarriers that limits the system data rate. Here, we establish an optical technique for carrier recovery harnessing large-gain stimulated Brillouin scattering (SBS) on a photonic chip for up to 116.82 Gbit·s$^{-1}$ Self-CO-OFDM signals, without requiring a separate LO. The narrow SBS linewidth allows for a record-breaking small carrier guardband of ~265 MHz in Self-CO-OFDM, resulting in higher capacity than benchmark self-coherent multi-carrier schemes. Chip-based SBS-self-coherent technology reveals comparable performance to state-of-the-art coherent optical receivers while relaxing the requirements of the DSP. In contrast to SBS processing on-fiber, our solution provides phase and polarization stability. Our demonstration develops a low-noise and frequency-preserving filter that synchronously regenerates a low-power narrowband optical tone, which could relax the requirements on very-high-order modulation signaling for future communication networks. The proposed hybrid carrier filtering-and-regeneration technique could be useful in long-baseline interferometry for precision optical timing or reconstructing a reference tone for quantum-state measurements.**

Modern fiber-optic coherent communications have triggered higher-order than 4-states advanced modulation formats such as 16–128-quadrature amplitude modulation (QAM), requiring sophisticated narrower linewidth (<100 kHz) local oscillators (LOs) compared to common distributed feedback lasers (>1 MHz) [1]. On the other hand, a significant frequency mismatch between the transmitter laser (carrier) and the receiver LO-laser occurs in coherent optical transmission systems that highly impacts the phase coherence producing phase noise. More specifically, due to the frequency mismatch of the transmitter laser and the LO, a carrier frequency offset (usually time varying) occurs and since the transmitter and LO lasers are not phase-locked, this necessitates a carrier phase recovery stage at the receiver. To cancel this frequency mismatch, an extra digital signal processing (DSP) functional block is traditionally employed which increases the DSP computational load and potentially both electronic processing power [2-4] and latency [5]. This is perceived to be critical for future-proof real-time machine-to-machine communications such as remote medicine, financial trading [5], cloud computing and the Internet-of-Things (IoT).

Whereas previous studies [2-4] tackle the Kerr-induced fiber nonlinearity which is critical for long-haul coherent optical communications, presently there is no solution to relax the stringent requirements of DSP units and advanced modulation formats in cost- and-energy-sensitive short-reach communications such as warehouse-scale data centers [6], access and metropolitan networks [7] (including 5G [8]), that have become coherent-centric to support capacities [9, 10] >100 Gbit·s$^{-1}$. Alternative approaches for practical short-reach coherent communications systems have been demonstrated: for instance, Kramer-Kronig systems [11, 12,] reduce the receiver optical complexity by utilizing a single photodiode, however, advanced DSP is still required.

Self-homodyne coherent technology [13-22] where the LO laser is transmitted together with the signal relaxes the requirements for DSP since eliminates the need for frequency offset compensation between transmitter laser and LO. Yet, for single-carrier modulation this technology sacrifices one of the two orthogonal polarizations of the fiber to fit an optical carrier that halves transmission capacity. For multi-carrier schemes such as self-coherent optical orthogonal frequency-division multiplexing (Self-CO-OFDM) [14, 15, 19, 20, 22], it typically requires a large carrier guardband (i.e. frequency gap that discards many OFDM middle subcarriers) for filtering-out the carrier that significantly limits signal capacity. Recent carrier recovery solutions employing injection-locked lasers [14, 15] and Fabry-Perot filters [20] have been reported but they lack the spectral resolution required for high spectral efficiency or require high-precision carrier-tracking analogue devices at the receiver.

In this paper, we propose to harness large-gain stimulated Brillouin scattering (SBS) [23-25] on a compact photonic chip for ultrahigh-resolution selective filtering for carrier recovery in high-capacity self-coherent optical signals without requiring a separate LO. SBS is processed on a chalcogenide chip to enable sufficient high gain, thus providing successful extraction and regeneration of the carrier. The ultrahigh-resolution SBS selective filtering of the received carrier increases signal capacity by reducing the carrier guardband between the carrier and the signal-band. We report that on-chip SBS demonstrates a high-performance practical ('black-box') "self-tracking" narrowband hybrid amplifier-and-filter without requiring analogue loops. Our demonstration merges for the first time the attractive features of chip-based SBS [23, 24] and self-coherent technology.

Previously reported fiber-based SBS-self-coherent systems [26, 27] have been limited to single-quadrature modulated formats to avoid excessive phase noise from the interaction of SBS with the processed fiber (typically of long length) that results resulting in polarization drifts and large propagation delays and in increased susceptibility to environmental fluctuations. In comparison, our chip-based approach is free from excessive carrier drifting providing phase and polarization stability associated with thermal and polarization drifts over the propagation length of the Brillouin gain medium, thus eliminating the need for high-precision carrier-tracking analogue devices at the receiver. Due to the narrow linewidth of SBS and the "self-referencing" nature of the proposed technique a record-breaking narrow carrier guardband of ~265 MHz for a 116.82 Gbit·s$^{-1}$ is achieved for self-coherent optical signals. Our approach reveals comparable performance to state-of-the-art coherent optical receivers without requiring additional digital techniques.

## 1. PROOF-OF-CONCEPT AND COMPARISON WITH BENCHMARK TECHNOOGIES

The principle of the compact photonic chip scheme for carrier recovery in high-capacity self-coherent optical signals is summarized in Fig. 1. For modulation, we used OFDM [28] which is constituted of MHz-bandwidth subcarriers, $M_s$. To this end, we established an *ad-hoc* SBS-based Self-CO-OFDM system as a test case for carrier recovery without requiring DSP carrier recovery. On-chip SBS-Self-CO-OFDM harnesses the unique capability of SBS to retrieve frequency, amplitude, and phase information from the extracted carrier without employing a separate LO. The narrow linewidth of SBS and the "self-referencing" nature of the proposed technique allows for a narrow carrier guardband of ~265 MHz (without considering the bandwidth allocated to the DC carrier), which is set in the middle of the OFDM signal to easily filter-out the received carrier. In contrast to benchmark Self-CO-OFDM schemes [13-20], our chip-based approach offers a compact stable integrated solution and enhanced signal capacity by reducing the frequency carrier guardband. The SBS filter is free from analogue carrier-locking lasers devices and electronics [14, 15, 20] due to the self-tracking nature of the filter formed by using a frequency-shifted copy of the received signal as the pump and has narrowband function (3-dB bandwidth of 10's of MHz) which is much tighter than traditional solutions employing injection-locked lasers [14, 15] and Fabry-Perot filters [20]. For our experiments the Self-CO-OFDM system was set at a nominal data rate, $S_k$, of 116.82 Gbit·s$^{-1}$ for 16-QAM and 54.95 Gbit·s$^{-1}$ for quaternary phase-shift keying (QPSK) using 127 generated subcarriers, $M_s$. Modulation was performed using an IQ-Mach–Zehnder modulator (IQ-MZM) for amplitude (I) and phase (Q) data subcarriers. It should be noted that using Alamouti coding of the signal [28] polarization independency is feasible without a polarizing element in place. While Alamouti coding causes 50% redundancy due to the replication of the transmitted symbols and requires a polarisation modulator in the transmitter, however, it does not require any additional DSP complexity, higher bandwidth or resolution of the DACs/ADCs. Furthermore, it enables to remove the polarisation beam splitter(s)/rotator(s), the 90° optical hybrid and two of the balanced photodetectors. The polarisation-independent single polarisation coherent receiver enabled by Alamouti coding retains many of the advantages of coherent detection including high power/bandwidth efficiency, linear optical field detection, and robustness to fibre impairments. Another option could be through aligning the carrier with the signal in a single polarization, or speculatively by combining the current scheme with polarization diverse or independent injection locking [15]. The OFDM signal was transmitted through 40 km standard single-mode fiber (SSMF) link, and afterwards, a 50:50 coupler was used at the receiver to separate the OFDM signal path from carrier extraction as depicted in Fig. 1(a). In the carrier extraction path, the SBS selects and amplifies the optical carrier, operating as a narrow-width optical band-pass self-tracking filter (OBPF) that follows a Lorentzian spectral profile as depicted in Fig. 1(b) (with a gain measured up to 20 dB for a chalcogenide chip alone using a pump-probe technique described in [23]. Note that in the carrier-recovery experiment, a lower gain was achieved. This is because the 'signal' power around the pump does not contribute to the SBS amplification). The self-tracking nature of the filter is realized by splitting the signal and using the frequency-shifted version of the signal as the SBS pump. So even if the laser drifts, the SBS pump will drift in the same manner, thus always providing SBS gain on the signal. At the photodetector, the beat between the LO and the signal will not drift in the RF domain, thus nullifying the need for frequency-offset compensation.

**(a)**

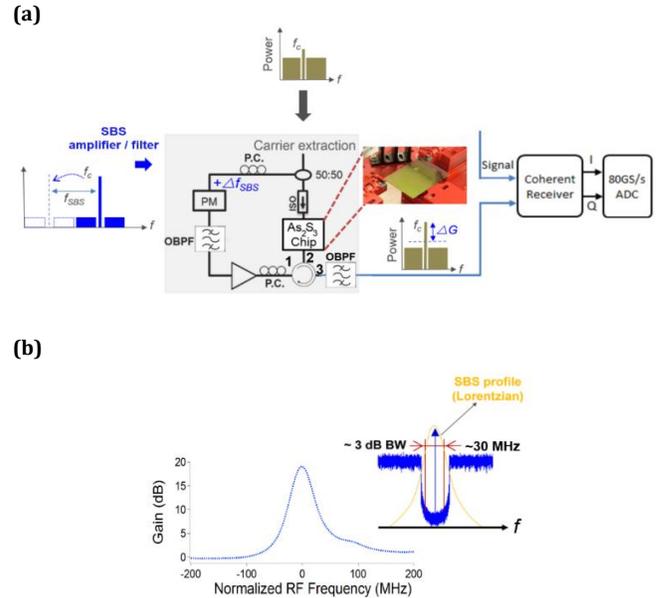

**(b)**

**Fig. 1:** (a) Receiver SBS-based Self-CO-OFDM. (b) Filter gain profile (Lorentzian) and illustration of the filtering process on an OFDM signal spectrum with carrier guardband of ~256 MHz. *P.C.: polarization controller, (O)BPF: (optical) band-pass filter, ISO: isolator; ADC: analogue-to-digital converter; OM: optical modulator, PM: phase modulator, As$_2$S$_3$: chalcogenide.*

The SBS gain bandwidth was measured to be only ~30 MHz. In the SBS process, another OBPF of 5 GHz bandwidth selected only the upper sideband which was amplified and counter-propagated with the other half of the received signal in the photonic chip. The Brillouin amplified carrier was extracted through port 3 of the circulator and the output of the SBS-pump was sent to an external OBPF of ~5 GHz bandwidth to primarily remove the back-reflected pump and it also aids in the increase of the optical carrier-to-signal ratio (OCSR) of the transmitted signal as shown in Fig. 3. This can be removed through optimized waveguide fabrication to minimize back-reflections. Finally, the recovered carrier was sent to the LO input of the coherent homodyne receiver which was connected to an oscilloscope, and the received data were processed offline using Matlab®. For our experiments, the Brillouin amplification was performed in a 24.0 cm long chalcogenide (As$_2$S$_3$) chip. In the first experiment, the proposed Self-CO-OFDM system was compared to a state-of-the-art CO-OFDM which employs DSP-based phase noise and frequency offset compensation. In the second experiment, chip-based SBS was compared to a Self-CO-OFDM system with SBS being realized in a 4.46 km-long SSMF. For the chip-based approach SBS gain is measured, when the chip is inserted into the signal path and is therefore the ON-OFF gain, measured to be 14 dB. The propagation loss is about 0.2 dB/cm. Characterization for similar chips has been provided in Ref. [30]. The total and individual OFDM $M_s$ subcarriers' bit-error-rate (BER) by error counting, Q-factor ($=20\log_{10}[\sqrt{2}erfc^{-1}(2BER)]$), and SBS gain were the crucial parameters under investigation. The received signal was also noise-loaded using an amplified spontaneous emission source to measure the Q-factor as a function of the optical signal-to-noise ratio (OSNR).

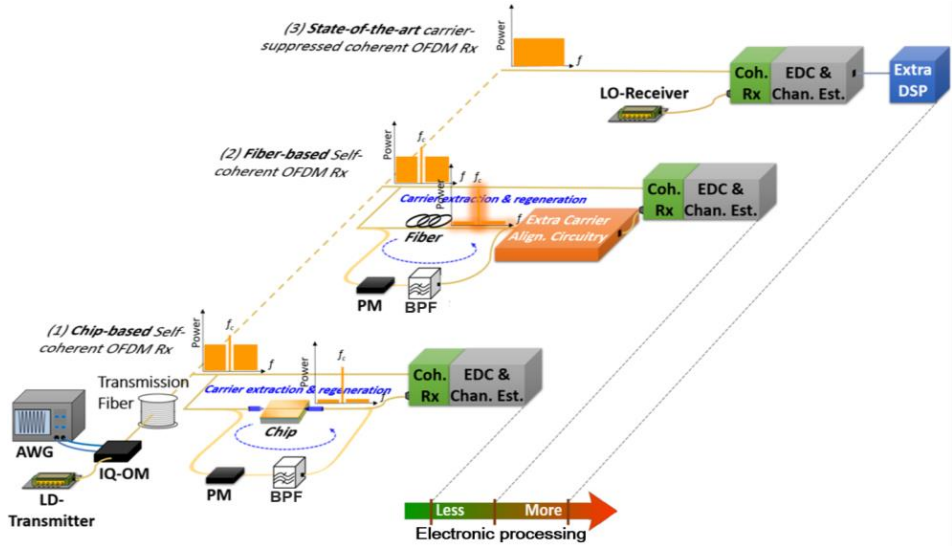

**Fig. 2:** Qualitative comparison of the developed *(1)* chip-based Self-CO-OFDM system with *(2)* SBS processed on a fiber, and *(3)* a state-of-the-art carrier-suppressed CO-OFDM. *AWG: arbitrary waveform generator, LD: laser diode, OM: optical modulator, PM: phase modulator, EDC: electronic dispersion compensation, LO: local oscillator, DSP: digital signal processing.*

A qualitative representation and comparison between the developed SBS carrier recovery chip for self-coherent detection, a fiber-based SBS processor, and a state-of-the-art carrier-suppressed coherent system with OFDM is illustrated in Fig. 2. In this block-diagram, we show that the proposed chip-based solution can essentially provide less electronic processing compared to the other two approaches. This occurs because, firstly, chip-based SBS can extract and regenerate the optical carrier from the coherent signal without employing frequency alignment circuitry at the receiver. Secondly, it only requires electronic dispersion compensation (EDC) and channel estimation at the receiver, eliminating the need of additional DSP for both frequency-offset correction and phase noise compensation.

## 2. METHODS

Below we provide more details about the experimental and simulation setups:

### A. Experimental setup

In Fig. 3, the Self-CO-OFDM experiment is depicted including the SBS-based 'black-box'. From the inset optical spectrum analyzer-1 (OSA-1) of Fig. 3, the OCSR was measured around −3 dB (at 5 MHz spectral resolution). It should be indicated that the optical signal power in our OSNR measurement is the combined signal and carrier power; hence, the ratio of the actual OSNR to the measured OSNR is 1+OCSR (using linear ratios). In Fig. 4, the experimental setup of the conventional 16-QAM CO-OFDM is depicted in which a LO is implemented at the receiver. The OCSR for the CO-OFDM was set at -22 dB (below and above which the system performance is degraded). In more detail, the baseband waveform samples were calculated offline based on a pseudo-random binary sequence of $2^{19}-1$. In the transmitter, an arbitrary waveform generator was used at a sampling rate, $r_s$, (frequency bandwidth) of 34 GHz to generate a continuous Self-CO-OFDM or conventional CO-OFDM baseband signal. The frequency spacing from center-to-center k-subcarrier was estimated at ~133 MHz. For the generation of the OFDM signal, a 256-inverse fast-Fourier transform (IFFT) size was implemented from which 128 subcarriers (1000 symbols per subcarrier) were employed as data and the rest were set to zero (highest 128 frequencies) in order to separate the OFDM baseband signal from the aliasing products generated at the output of the digital-to-analogue converters (DACs). The modulation format was either QPSK or 16-QAM (with the total number of binary bits, $n_k$, to be 2 and 4, respectively) and only 1 subcarrier around the direct current (DC) component was padded with zero (carrier guardband) to enable a spectral guardband within OFDM subcarriers in Self-CO-OFDM and allow the insertion of an optical carrier (i.e. [128−1] $M_s$ subcarriers). A cyclic prefix overhead of 10% was inserted to increase inter-symbol interference tolerance. A 3% of the signal band was sacrificed for channel estimation (pilot-aided) for both QPSK and 16-QAM and no carrier frequency offset was required. Therefore, the nominal data rates, $S_k$, resulted in 54.949 Gbit·s⁻¹ (QPSK) and 116.82 Gbit·s⁻¹ (16-QAM) for Self-CO-OFDM.

In optical domain, a tunable external cavity laser with a linewidth of ~100 kHz (this value is lower than the SBS gain bandwidth allowing stable operation condition) was used at the transmitter to generate a continuous wave signal at ~1,550 nm (signal pump, ωp) that was subsequently modulated with the Self-CO-OFDM signal by an IQ-MZM, biased such that the carrier was not totally suppressed. We should note that since the modulator can operate at a constant temperature using a TEC, then there is no need for feedback to a bias controller, as bias remains stable. The launched optical power for both on-chip and SMF-based Self-CO-OFDM systems and CO-OFDM was set at 0 dBm. After transmission through a 40 km of SSMF link, a noise source was placed using an EDFA to adjust the OSNR at a desired value. At the receiver, a 50:50 coupler was used to separate the OFDM signal path from the carrier extraction path. In the carrier extraction path, the SBS pump extracts and amplifies the carrier where we counter-propagated the frequency-shifted signal in the SBS medium. Another 50:50 coupler was used to split the light in the carrier extraction path with half of the light being modulated by a phase modulator (PM) to shift the incoming light to by the Brillouin stokes shift. In fiber, the Brillouin shift is 10.86 GHz

[25], while for the chalcogenide waveguide on chip, the shift is 7.7 GHz [23]. The output of the SBS pump was sent to a combined EDFA with variable optical attenuator (VOA), as a safety measure to protect the coherent receiver from any fluctuations in carrier power (~0.5–0.7 dB) due to SBS gain variations from polarization or other environmental effects. This was subsequently incident on a 90° optical hybrid and emulated as LO. The polarization of the extracted optical carrier and Self-CO-OFDM signal were aligned with polarization controller in the Self-CO-OFDM signal path before detection. Finally, the sampling speed of the digital storage oscilloscope was 80 GS/s in which the offline receiver DSP was executed in Matlab.

For the case of conventional CO-OFDM identical parameters with Self-CO-OFDM were adopted with the exceptions of employing DSP-based frequency offset compensation with the assistance of pilot subcarriers and not setting to zero any data-subcarrier (thus generating 128 subcarriers). In conventional CO-OFDM the optical carrier was suppressed (-22 dB of OCSR). Moreover, an external cavity laser with 100 kHz linewidth was used as LO, while the launched optical power of CO-OFDM was also set at 0 dBm. 3% (optimum) of the signal band was sacrificed for channel estimation (pilot-aided) in CO-OFDM for both QPSK and 16-QAM. However, in contrast to Self-CO-OFDM the nominal $S_k$ dropped to 49.261 and 107.22 Gbit·s-1 for QPSK and 16-QAM, respectively: this occurred due to the sacrifice of 5% (QPSK) and 8% (16-QAM) of generated subcarriers for pilot-assisted frequency offset compensation corresponding to 6 and 10 pilot subcarriers (optimum), respectively.

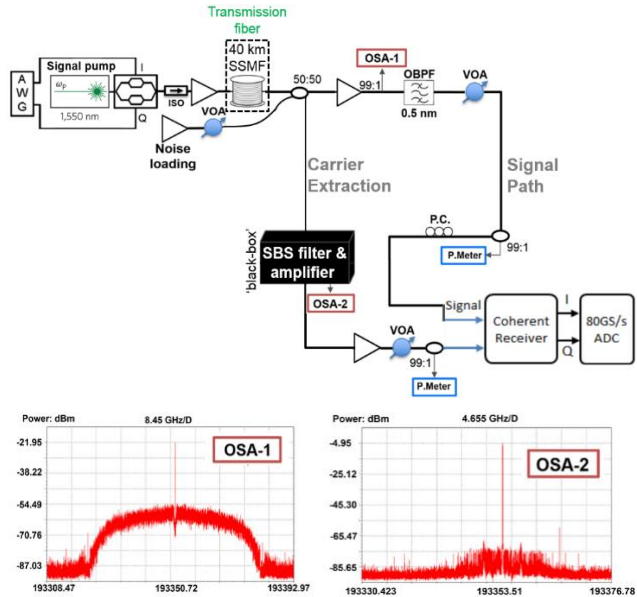

**Fig. 3:** SBS-based Self-CO-OFDM experimental setup block diagram. Insets: Optical spectrums from transmitter-side (OSA-1) and after carrier selection (OSA-2). OSA: optical spectrum analyzer; VOA: variable optical attenuator, P. Meter: power meter.

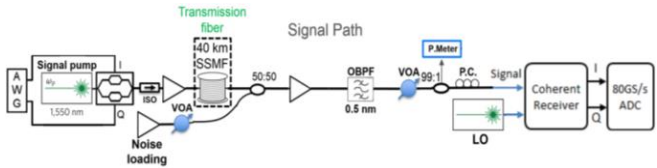

**Fig. 4:** Experimental conventional CO-OFDM block diagram that includes the transmission-link and noise loading.

### B. Waveguide fabrication

Chalcogenide ($As_2S_3$) thin films were deposited via thermal evaporation on <100> oriented 100 mm thermal oxide Silicon wafers and a total film thickness of 930 nm was achieved. A ~100 nm layer of SU8 was coated on the wafer before annealing at 130°C for 24h. Through projection lithography waveguides with nominal widths of 2.2 μm, 2.4 μm and 2.6 μm were patterned on the $As_2S_3$ layer. The waveguides were designed in spiral coils with waveguide length of 23.7 cm in a physical length of 2 cm. After ICP - inductively coupling plasma - etching an etch depth of 330 nm was targeted and an additional layer of $SiO_2$ was overclad for enhanced acoustic confinement.

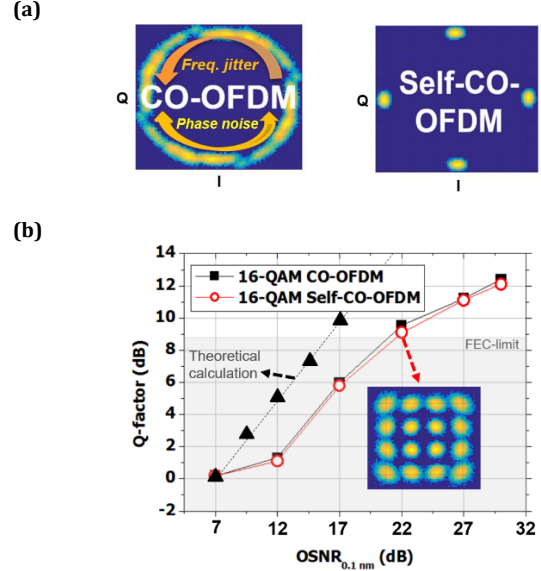

**Fig. 5:** (a) Comparison of received quaternary phase-shift keying (QPSK) constellation diagrams. (b) Comparison at 40 km: Quality (Q)-factor vs. optical signal-to-noise ratio (OSNR) for experimental 16-quadrature amplitude modulation (16-QAM). Black triangles refer to the theoretical calculation. The SBS gain and optical carrier-to-signal ratio (OCSR) for the self-coherent system were fixed at 14 dB and −3 dB, respectively. The launched optical power for both systems was fixed at 0 dBm. *FEC: forward-error-correction.*

### C. Simulation set-up and parameters

The simulation parameters for the developed Self-CO-OFDM system including the 40 km optical link were identical to the experimental setup, being implemented in a Matlab/Virtual Photonics Inc. (VPI)-transmission-Maker® co-simulated environment (electrical domain in Matlab and optical components with SSMF in VPI). The SBS noise was modelled as additive-white Gaussian noise (AWGN). The Self-CO-OFDM modelling was based on the selective amplification of the carrier with a Lorentzian gain profile (using the optimum SBS gain value) followed by AWGN, whose spectral density was determined from the measured noise figure of the Brillouin amplification process. For the in-line optical amplification, an EDFA was adopted having 8 dB gain and 5.5 dB of NF. The EDFA noise was also modelled as AWGN. The SSMF for single-polarization transmission was modelled using the pseudo-spectral split-step Fourier method which solves the nonlinear Schrödinger equation [2]. The adopted SSMF parameters in this work are the following: fiber nonlinear Kerr parameter, chromatic dispersion, chromatic dispersion-slope, fiber loss, and polarization-mode dispersion coefficient of 1.1 W$^{-1}$ km$^{-1}$, 16 ps nm$^{-1}$ km$^{-1}$, 0.06 ps km$^{-1}$ (nm$^2$)$^{-1}$, 0.2 dB km$^{-1}$ and 0.1 ps (km$^{0.5}$)$^{-1}$, respectively. Pilot-aided phase noise compensation was implemented for conventional CO-OFDM which sacrifices useful signal bandwidth. This is derived from the signal capacity $R_{signal}$ expressions [31] for coherent signals:

$$T_S = \frac{2M_S(1+p)}{r_s} \quad (1)$$

$$R_{signal} = \sum_{k=1}^{M_S} S_k = \frac{\sum_{k=1}^{M_S} n_k}{T_S} = \frac{r_s \sum_{k=1}^{M_S} n_k}{M_S(1+p)} \quad (2)$$

where $S_k$ is the signal bit-rate corresponding to the k-th subcarrier, $M_s$ is the number of subcarriers, $n_k$ is the total number of binary bits conveyed by the $k^{th}$ OFDM subcarrier within one symbol period $T_s$, $r_s$ is the sampling rate, and $p$ includes the cyclic prefix length to increase the inter-symbol interference tolerance and the total pilot-tune period (related to the pilot-aided symbols for frequency-offset estimation).

## 3. RESULTS

### A. Comparison to a state-of-the-art coherent optical system

In Fig. 5(a), the performance benefit by the on-chip SBS-based Self-CO-OFDM is illustrated using QPSK modulation for simplicity. The OSNR was set at 27 dB to isolate the transmitted fiber-induced effects at 40 km from random optical amplification noise (amplified spontaneous emission). In this illustration, digital-based carrier recovery has been neglected for both Self-CO-OFDM and CO-OFDM. It is clearly indicated that the received symbols in the QPSK constellation diagram of CO-OFDM are highly distorted causing "phase rotation" in contrast to our chip-based approach. This example shows the necessary requirement of state-of-the-art coherent systems for extra DSP-based phase noise and frequency offset compensation. In Fig. 5(b), the performance of conventional 16-QAM CO-OFDM is compared to a chip-based 16-QAM Self-CO-OFDM at 40 km of transmission using an SBS gain and launched optical power of 14 dB and 0 dBm, respectively. This gain enables performance above the forward-error-correction (FEC)-limit according to a targeted BER [32] of $3.3 \times 10^{-3}$. For Self-CO-OFDM, no phase or frequency offset correction was applied at the receiver-side DSP, therefore we investigated a truly self-coherent system. The OCSR measured as the ratio of the carrier to the level of the signal power from the optical spectrum analyzer (spectral power density), was set at optimum $\sim -3$ dB in Self-CO-OFDM and at $-22$ dB (suppressed carrier) in CO-OFDM. It should be noted for the case of fiber-based Self-CO-OFDM that when the OCSR is very large, the SNRs on neighbour subcarriers of the carrier are reduced compared to the chip because the increased fiber length results in susceptibility to environmental variations thereby causing performance penalties, i.e. an increase of both amplitude and phase noise. When the OCSR is too low, the carrier cannot be effectively recovered since the receiver blurs the carrier with the neighbour subcarriers and also optical amplification noise limits the effectiveness of phase noise compensation. The results depicted in Fig. 5(b) indicate there is no measurable OSNR penalty when including an enhanced carrier. This specifies that the system performance of our method is completely comparable with an optimized coherent optical system, revealing that the recovered carrier with the SBS pump operates similarly to an external cavity laser for practical levels of signal noise. On the other hand, 16-QAM Self-CO-OFDM outperforms conventional CO-OFDM in terms of signal capacity by 9.6 Gbit·s$^{-1}$ since pilot-assisted frequency compensation is not necessary.

Compared to the 16-QAM theoretical curve [33], our required OSNR shows an implementation penalty of ~4.7 dB at the FEC-limit. We attribute this implementation penalty to bandwidth limitations of the RF components and the modulator; the fact that we did not pre-equalize the drive signals and the additional noise from the coherent receiver, arbitrary waveform generator and digital storage oscilloscope. Moreover, since pilot-subcarrier aided phase noise compensation [34] is implemented in CO-OFDM sacrificing 8% of the signal, the data rate, $S_k$, of 16-QAM Self-CO-OFDM is enhanced by 9.6 Gbit·s$^{-1}$ following Eqs. (1), (2). Pilot-aided algorithms in CO-OFDM can virtually compensate any residual phase noise from carrier drifting (measured up to 1 MHz). On the contrary, implementing fully-blind phase noise estimation in CO-OFDM is very challenging without considering complex differential bit encoding due to CO-OFDM's long symbol duration and large carrier offset [35]. Finally, it is worth noting that a similar transmission performance is also anticipated from benchmark single-carrier modulation schemes such as Nyquist wavelength division multiplexing [36].

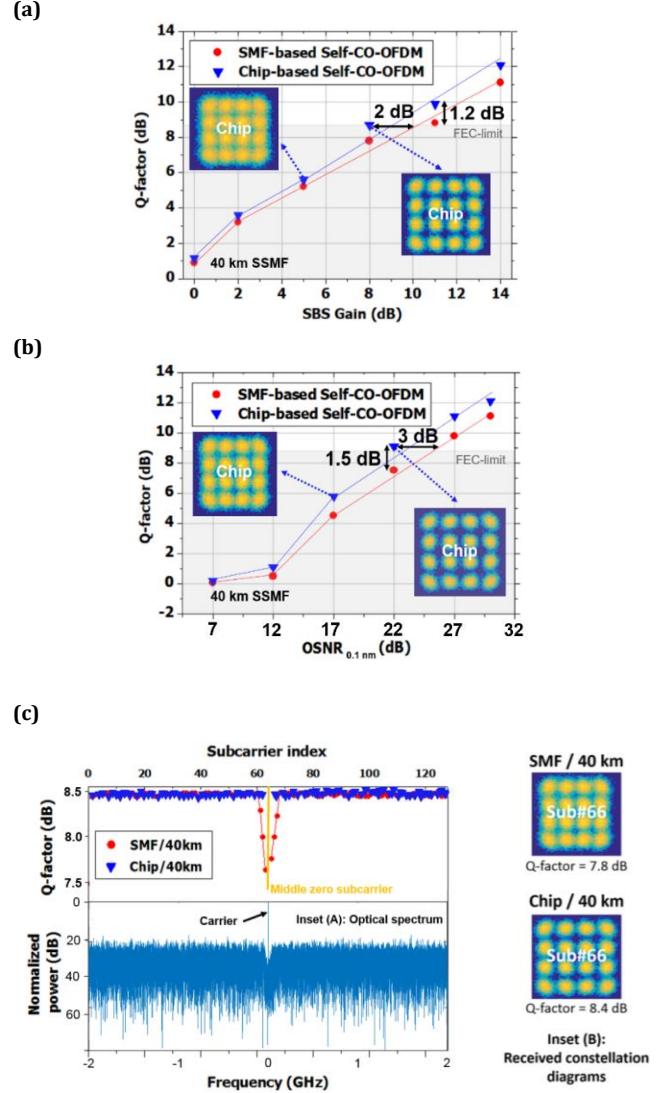

**Fig. 6:** Comparison of on-chip and fiber-based SBS for 16-QAM Self-CO-OFDM. (a) Q-factor vs. SBS gain with fixed OSNR at 38 dB and optimum OCSR at –3 dB. (b) Q-factor vs. OSNR for best SBS-gain at 14 dB. (c) Subcarrier index (127 subcarriers) vs. Q-factor for on-chip and single-mode fiber (SMF) based SBS with OSNR and gain fixed at 38 and 8 dB, respectively. Inset (A): Received optical spectrum with a resolution of 5 MHz after 40 km of SMF transmission. Inset (B): Example of received constellation diagrams and Q-factors for subcarrier#66.

### B. Comparison to a fiber-based SBS-self-coherent OFDM

The chip-based carrier recovery is compared to a fiber-based approach to SBS processing using 4.46 km of SSMF, highlighting the significant drifting in the laser frequency and phase. The reason for using the SSMF was to ensure a single SBS peak, which is critical for this experiment. A fiber with multiple SBS peaks would amplify signals out-of-band and would also increase the required SBS pump power. The length of the fiber was chosen to: a) achieve high gains with low pump powers while also limiting the added SBS noise, and b) be consistent with other works reported on SBS-based processing [27]. For this demonstration, the impact of the SBS gain and OSNR on the 16-QAM Self-CO-OFDM performance is investigated in Fig. 6. The OCSR was maintained at the optimum value of –3 dB. It should be noted that the fiber-based Brillouin gain is also in the saturation regime similarly to Ref. [27]. In Fig. 6(a), the OSNR was set at a high value of 38 dB to ensure that amplified spontaneous emission noise will not affect our system, keeping the Q-factor within the FEC-limit. For low SBS gain, there is no clear distinction between the on-chip and fiber-based SBS. However, for a high gain (> 6 dB) on-chip SBS outperforms up to ~1.2 dB in Q-factor, while at the FEC-limit it reduces the maximum required SBS gain by ~2 dB. In Fig. 6(b) the SBS-gain was fixed at 14 dB to ensure signal quality well-above the FEC-limit, validating perfect carrier recovery on the OSNR evolution. Likewise, for a low OSNR nearly identical performance is noticeable between chip- and fiber-based SBS. Though, for high OSNR the on-chip SBS outperforms up to 1.5 dB in Q-factor and at the FEC-limit it extends the required OSNR by 3 dB.

In conclusion, results here reveal that SBS on a chip outperforms to a fiber-based process at high gains. This reveals for the first time, the technological advantage of using photonic chips over optical fibers beyond the well-established arguments of size, weight, power and cost (SWAP-C). This indicates that the photonic chip is a necessary component for elimination of excessive carrier drifting and phase noise, resulting in the enhancement of a few subcarriers' SNR that are located around the carrier guardband. On the other hand, for low SBS-gain both systems present similar performance since they are dominated by Gaussian noise. The chip-based improved performance is corroborated in Fig. 6(c), where the individual subcarrier Q-factor is plotted for 16-QAM Self-CO-OFDM at 8 dB of SBS gain. To that end, we identify that the Q-factors of the center subcarriers are improved when chip-based SBS is applied due to the cancellation of the residual phase noise. This is explained as follows: in a fiber, SBS suffers more drifts because it is longer and exposed more to thermal drifts, at least relative to the chip that is integrated [37]. This results the SBS detuning to shift the gain in a fiber thus boosting the power of the nearest neighbour subcarriers of the carrier, producing unwanted signal-signal beating. These subcarriers are located around the frequency carrier guardband of the OFDM signal, as shown in the received optical spectrum (5 MHz resolution) of inset (A) of Fig. 6(c) related to 40 km transmission. An example of the received constellation diagrams and Q-factors is illustrated in inset (B) for subcarrier#66, revealing an improved Q-factor of 0.6 dB using on-chip SBS.

### C. Chip-based Brillouin noise simulation analysis and envisioned chip solution

In Fig. 7, the impact of SBS noise on the performance of 16-QAM SCO-OFDM is simulated at 40 km transmission and compared with experimental traces. The simulated results are related to an ideally-modelled Self-CO-OFDM in which frequency drifting was excluded and the OCSR and gain were set similarly to our experiment at –3 dB and 14 dB, respectively. Results are therefore related to chip-based SBS which has shown no sign of carrier drifting. More details of the modelled Self-CO-OFDM are provided in Methods. As depicted in Fig. 7, the emulated SBS noise figure (NF) is estimated roughly at maximum 5 dB considering there is some implementation penalty. Such levels of noise figures can be expected for the strong saturation regime of SBS [38-40]. Yet, in comparison to the case of excluding SBS noise, only ~0.95 dB of Q-factor degradation is observed at 30 dB of OSNR (FEC-limit). Hence, the simulation analysis shows that the additional noise added to the carrier is equivalent to that added by a 5 dB NF amplifier, indicating that the SBS amplification has a NF comparable with other optical amplification schemes. However, it should be noted that this is a topic of research that warrants further investigation.

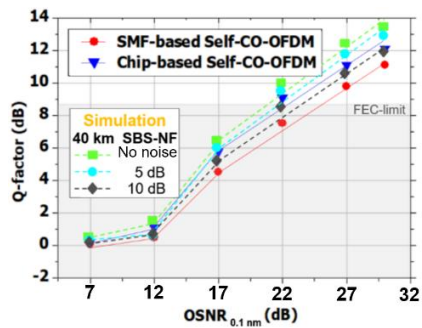

**Fig. 7:** SBS noise limit by simulation analysis. Q-factor vs. OSNR for 16-QAM Self-CO-OFDM with optimum OCSR at –3 dB and best SBS gain at 14 dB (according to Fig. 4). Simulated results for ideal Self-CO-OFDM (i.e. no frequency drifting) for different SBS noise figure (NF) are related to 40 km transmission at a launched optical power of 0 dBm.

Finally, we propose that full integration of an SBS-self-coherent receiver on a compact photonic chip can potentially further reduce power consumption, as well as footprint and weight [40]. Recent progress in developing photonic-chip based SBS devices offer potential for integrating optical waveguides that exhibit high SBS gain, together with critical components such as optical modulators and photodetectors, in a single photonic chip will enhance system stability. An illustration of such envisioned integrated carrier recovery chip combining the Brillouin processor in chalcogenide, circuits in silicon and modulator, filter, amplifier, and the coherent receiver in indium phosphide is shown in Fig. 8. To this end, a first step in hybrid integration of Silicon and chalcogenide has been taken, where 18.5 dB Brillouin gain was achieved in a chalcogenide footprint [41] of only 100 $\mu$m × 4 mm. A total pump power of ~23 dBm was used to achieve 14 dB of SBS gain. The on-chip power was 19 dBm due to the ~4 dB/facet coupling losses. With optimization of the waveguide design, the propagation losses can be reduced even further decreasing the required pump power to ~17 dBm, power levels that are compatible with on-chip amplifiers as shown in Fig. 8. In the current embodiment, external optical amplifiers are required for boosting the pump powers.

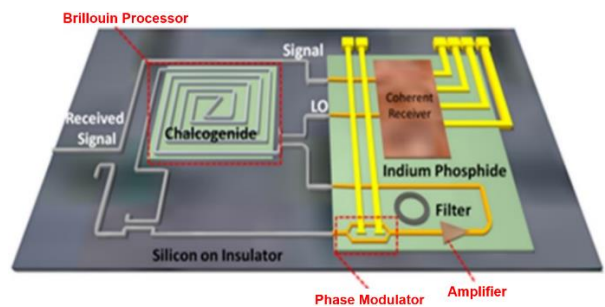

**Fig. 8:** Envisioned chip-based solution: an integrated photonic chip with carrier recovery and self-coherent detection. In this chip, the chalcogenide ($As_2S_3$) waveguide [23, 24] is used as the phononic processor for providing narrowband Brillouin gain, silicon components to support functional circuits, and indium phosphide for active devices including detectors.

### 4. CONCLUSION

In this work, we reported a novel carrier recovery technique for high-capacity self-coherent optical signals harnessing on-chip SBS that functions as a self-tracking filter. Our demonstration developed a stable and robust all-optical hybrid amplifier-and-filter device that enables record-breaking narrowband of ~265 MHz-bandwidth (~485 MHz smaller carrier guardband than Ref. [15] for 16-QAM subcarriers) carrier extraction and regeneration in Self-CO-OFDM; without requiring analogue loops neither a separate LO-laser. The proposed low-complex system was tested in cost-and-energy sensitive short-reach telecommunications for signal capacities of up to > 100 Gbit·s$^{-1}$ using high-order modulation formats. However, since it has been shown in some previous self-coherent demonstrations [13, 42] that nonlinear distortions can be tolerated, it is anticipated our approach to be considered for longer distances. In contrast to a fiber-based SBS-self-coherent system, our approach provided both phase and polarization stability by neutralizing residual carrier frequency drifting, without requiring sophisticated carrier-tracking analogue devices at the receiver. In comparison to state-of-the-art coherent optical systems, SBS-self-coherent technology discarded the need for digital-based phase noise and frequency offset compensation, thus relaxing the receiver complexity by removing a DSP functional block without sacrificing transmission performance. To this end, pilot-subcarrier aided phase noise compensation was not required for self-coherent OFDM that wastes useful bandwidth. Moreover, as modulation complexity increases, the computational cost of phase estimation in particular may be increased significantly. Hence, since our approach is fundamentally modulation format independent, it may afford a route to power savings in future systems. In comparison to Stokes receivers [43] our system eliminates the need of two extra photodetectors and relaxes the DSP complexity. On the other hand, as future-proof coherent systems will trigger higher-order than 16-states modulation formats (e.g. 32-QAM–128-QAM), requiring more complex DSP and sophisticated narrower linewidth LOs (<100 kHz) compared to common distributed feedback lasers (>1 MHz), our technique may relax requirements on very-high-order QAM signaling. Our technique is also transparent to higher spectral-efficient modulation formats (Fast-OFDM) [44] and non-orthogonal signal modulation techniques [45]. It should be noted that while the required SBS gain of benchmark fiber-based SBS-self-coherent systems is limited by the modulation format order, being less attractive for higher-order QAM signals compared to homodyne schemes, here, the large SBS gain (52 dB) provided by a chalcogenide chip [23] could potentially support up to 64-QAM [1] and above. Since QAM formats are extremely sensitive to phase and intensity noise [24, 46], our experiments confirmed the modest noise added by the narrowband SBS amplifier in a highly saturated regime. It has been shown that noise figure reduces significantly to values comparable to that of an Erbium-doped fiber amplifier (EDFA) under strong saturation [38]. The Brillouin amplifier with such low noise figure operating over a ~30 MHz bandwidth was the key to increasing the carrier power with respect to the carrier noise without introducing significant noise, therefore enabling successful modulation and demodulation of phase-sensitive signals.

We also believe that the realization of multiple functional devices in a compact photonic chip such as optical modulators and photodetectors could also enhance energy-efficiency and system stability — for example, thermal stabilization is easier to achieve for devices integrated on a single chip. On the other hand, the pump power for the SBS gain is close to 20 dBm which is comparable to the optical power used for LOs in standard coherent receivers. As such, the pump amplifier in our set-up should not consume much more power than a LO-laser. On-chip filtering can be low loss, and so not induce a high-power overhead. While the discrete modulators we use require significant power, on-chip modulators can be made to be very low power (e.g. using capacitive technology [47]). The predominant assumption for photonic integration is that this can reduce cost of relatively complex optical sub-systems, through mass-production of the sub-system itself.

Chip-based SBS-self-coherent optical transport systems are expected to find a wide range of applications to meet the ever-increasing demand of capacity upgrade and cost reduction in future optical networks. However, the proposed hybrid amplifier-filter device can potentially be employed for carrier recovery in other applications – picking out a narrow, "pure" optical tone with high-fidelity, could be useful in precision optical timing over very-long-baseline interferometry (VLBI) which may disclose important findings in fundamental physics and in the search for dark matter [48]. In VLBI, the correlation of signals from distant antennas starts with down-conversion and sampling at each telescope that is obtained using a LO, whose frequency is referenced to an atomic clock [48]. The spectral purity of the clock is of paramount importance, as in multiplication chains the phase noise is significantly increased, and hence, the atomic clock must exhibit an excellent long-term stability: a loss in the coherence between clocks used in different antennas results in fading of the interferometer fringes. Moreover, our proposed SBS-based device could be used for phase reference sharing in continuous-variable quantum key distribution [49]. Relying on a self-coherent approach, where in this case the phase reference information and quantum information are coherently obtained from a single optical wavefront, very-high phase noise can be tolerated [49]. Finally, we assume our solution can work for CDMA-based secure optical systems and other application scenarios with OFDM modulation and even machine learning [50-60].

**Funding**. Australian Research Council (ARC) through DECRA Fellowship (DE170100585, DE150101535), Laureate Fellowship (FL120100029), and Center of Excellence CUDOS (CE110001018).

**Acknowledgment**. We acknowledge the ACT node of the Australian National Fabrication Facility.